\documentclass[12pt]{iopart}
\usepackage{graphicx}

\newcounter{parentequation}

\def\ltsima{$\; \buildrel < \over \sim \;$}
\def\gtsima{$\; \buildrel > \over \sim \;$}
\def\simlt{\lower.5ex\hbox{\ltsima}}
\def\simgt{\lower.5ex\hbox{\gtsima}}
\def\planck{\it Planck\rm}
\def\plancks{\it Planck \rm}
\def\etal{{\it et al.}\rm}

\def\pmb#1{\setbox0=\hbox{#1}%
    \kern-.025em\copy0\kern-\wd0
    \kern.05em\copy0\kern-\wd0
    \kern-.025em\raise.0433em\box0}
\begin{document}

\title[B-mode detection with Planck]{B-mode Detection with an Extended
  Planck Mission}

\author{G. Efstathiou{\footnote[1]{Author to
whom correspondence should be addressed (gpe@ast.cam.ac.uk).}} and S. Gratton}

\address{ Kavli Institute for Cosmology Cambridge and 
Institute of Astronomy, Madingley Road, Cambridge, CB3 OHA.}

\begin{abstract}
  The \plancks satellite has a nominal mission lifetime of $14$ months
  allowing two complete surveys of the sky. Here we investigate the
  potential of an extended \plancks mission of four sky surveys to
  constrain primordial $B$-mode anisotropies in the presence of
  dominant Galactic polarized foreground emission. An extended
  \plancks mission is capable of powerful constraints on primordial
  $B$-modes at low multipoles, which cannot be probed by ground based
  or sub-orbital experiments. A tensor-scalar ratio of $r=0.05$ can be
  detected at a high significance level by an extended \plancks
  mission and it should be possible to set a $95\%$ upper limit of $r
  \simlt 0.03$ if the tensor-scalar ratio is vanishingly
  small. Furthermore, extending the \plancks mission to four sky
  surveys offers better control of polarized Galactic dust emission,
  since the $217$ GHz frequency band can be used as an effective dust
  template in addition to the $353$ GHz channel.

\medskip


\end{abstract}




\section{Introduction}

High precision experiments, most notably WMAP, have demonstrated that
the acoustic peak structure of the cosmic microwave background (CMB)
agrees extremely well with the predictions of simple inflationary
models $[1]$.  Nevertheless, relatively little can be inferred from
current observations about the detailed dynamics of inflation and some
cosmologists have questioned whether inflation actually took
place $[2]$.

The detection of a primordial $B$-mode polarization pattern in the CMB
would provide one of the strongest pieces of evidence that the
Universe experienced an inflationary phase $[3]$. Tensor perturbations
generated during inflation source $E$ and $B$-mode polarization
anisotropies of roughly equal magnitude, whereas scalar perturbations
produce only $E$-modes.  A detection of a $B$-mode anisotropy would
therefore confirm the existence of a background of stochastic
gravitational waves generated during inflation and would fix the
energy scale of inflation via
\begin{equation}
 V^{1/4} \approx 3.3 \times 10^{16} r^{1/4}  \; {\rm GeV} \label{I1}
\end{equation}
[4], where $r$ is the relative amplitude of the tensor and scalar 
power spectra (defined as in [5]). At present, the limits on $r$ are
relatively poor. Direct upper limits on the $B$-mode power spectrum
from the WMAP 5-year data set a limit of $r \simlt 10$ $[1]$. Indirect upper
limits, derived by combining  WMAP temperature and $E$-mode data with a variety
of other astrophysical data sets,  yield 95\% limits of $r < 0.22$ if
the scalar spectal index is set to a constant, and $r < 0.55$ is the scalar
spectral index is allowed to `run' ($dn_s/ d{\rm ln} k \ne 0$). 

To improve on these limits,  a number of sensitive ground
based/sub-orbital polarization experiments are either taking data or
under construction $[6]$.  The most sensitive of these experiments aim
to detect (or set limits) on a $B$-mode polarization pattern if $r$ is
as low as $\sim 10^{-2}$. In addition, groups in Europe and the USA
have considered designs for a future $B$-mode optimised space satellite
$[7]$.

The \plancks satellite is scheduled for launch in April
2009\footnote{For a description of \plancks and its science case, see
  $[8]$.}. \plancks has polarization sensitivity in $7$ channels over
the frequency range $30$-$353$ GHz.  However, \plancks was designed to
make high resolution ($\sim 5^\prime$) sensitive measurements of the
temperature and $E$-mode anisotropies. It was not optimised to detect
a large-angle $B$-mode polarization pattern. As a consequence,
\plancks lacks sensitivity compared to $B$-mode optimised experiments,
particularly those that use large bolometer arrays as
detectors. Nevertheless, there are several reasons to investigate
\planck's capability for $B$-mode detection in some detail:

\smallskip

\noindent 
[1] As a space experiment \plancks will operate in a stable
environment and covers a wide frequency range, offering control of
polarized synchrotron and dust foregrounds which will dominate over
any primordial $B$-mode signal.

\smallskip

\noindent 
[2] \plancks will scan the entire sky and is the only
experiment on the horizon which can probe polarization anisotropies at
low multipoles $\ell \simlt 10$.  \plancks therefore provides a useful
complement to ground based/sub-orbital experiments that probe higher
multipoles.

\smallskip 

\noindent
[3] For  inflation with a power law potential, $V(\phi) \propto \phi^\alpha$,
the scalar spectral index, $n_s$, and tensor-scalar ratio are approximately 
\begin{equation}
  n_s \approx 1 - {2 +\alpha \over 2N}, 
\qquad  r \approx { 4 \alpha \over N},  
\qquad {\it i.e.} \;\; r \approx 8(1- n_s) {\alpha \over \alpha + 2} 
\label{I2}
\end{equation}
(see {\it e.g.}\ $[9]$ for a review), where $N$ is the number of inflationary
e-folds between the time that CMB scales crossed the `horizon' and the
time that inflation ends. There are indications from WMAP and CMB
experiments probing higher multipoles for a small tilt in the scalar
spectral index $n_s \sim 0.97$ $[1]$, $[10]$. If we take this tilt
seriously, then the last of these equations suggests $r \sim 0.1$ for
any $\alpha$ of order unity. For example, for $N \approx 60$ $[11]$,
the quadratic potential gives $n_s \approx 0.97$, $r \approx
0.13$. This sets an interesting target for \plancks and other $B$-mode
experiments. `High field' inflation models with substantially lower
values of $r$ require special fine tuning (though phenomenological
models of this sort can certainly be constructed $[12]$).

Measuring inflationary $B$-modes on large angular scales is a
formidably difficult problem. For $r\sim 0.1$, the primordial B-mode
has an {\it rms} amplitude of of only $\sim 0.06$ $\mu$K on $7^\circ$ scales. The
signal is therefore small, but in addition, Galactic polarized
emission is expected to be $10$ to $20$ times higher over the high
Galactic latitude sky. The detection of a large-angular scale $B$-mode
signal therefore requires polarized foreground subtraction to an
accuracy of a few percent or better. In a recent paper $[13]$, we
presented a simple template-fitting scheme for modelling the
polarization likelihood from multifrequency data. The scheme was
tested using the Planck Sky Model (PSM) $[14]$ for polarized
foregrounds and found to work well for the signal-to-noise expected
for the nominal \plancks mission  which allows two full
sky surveys over $14$ months. The lifetime of the High Frequency
Instrument (covering the most sensitive CMB frequencies) on \plancks
is determined by the capacity of the $^3{\rm He}$ and $^4{\rm He}$
storage tanks that feed the $0.1$K dilution refrigerator. The precise
lifetime will depend on the operating conditions in-flight, but should
be sufficient to provide 4 full sky surveys with some considerable
margin.

The purpose of this paper is to analyse carefully the constraints on
primordial $B$-mode anisotropies that might be achieved with an
extended \plancks mission,  taking into account Galactic polarized
foregrounds. We will show that an extended \plancks mission offers
greater scope for testing the subtraction of polarized dust emission, 
and should be capable of detecting a tensor-scalar ratio of $r \approx
0.05$ at a high significance level. If the tensor-scalar ratio is 
very small ($ r \ll 0.05$), an extended \plancks mission should be able
to set a direct 95\% upper limit of $r \sim 0.03$. An extended
\plancks mission is therefore competitive with the most ambitious
of the polarization optimised ground-based/sub-orbital experiments.

\section{B-mode detection with Planck}

\subsection{Likelihood analysis of multifrequency data with foregrounds}

We first outline the procedure described in $[13]$. We designate
a subset of the frequency channels as `CMB sensitive' channels and
construct a data vector, $x_i$ (where $i$ is the index number of the
$3N_p$ pixels defining the map) as the inverse noise variance weighted
sum of the temperature $T$, and Stokes parameters $Q$ and $U$, over
these channels. The covariance matrix of this vector is written as
\begin{equation}
  \langle x_i x_j \rangle = S_{ij}+ \Phi_{ij} + N_{ij}, \label{B1}
\end{equation}
where $S_{ij}$ is the `signal' covariance matrix (primordial CMB), $\Phi_{ij}$
is the covariance matrix of the residual foreground contamination and $N_{ij}$
is the instrumental noise covariance matrix. To remove foregrounds, we designate
$N_F$ channels as templates, $F^k_i$, where the superscript denotes frequency,
and we construct the data vector
\begin{equation}
  Y_i = x_i - F_i^k \beta^k_i, \qquad (\beta^k_i =\beta^k_{(T, Q, U)}, 
\quad {\rm if} \ i \equiv (T, Q, U)). \label{B2}
\end{equation}
If the templates remove the foregrounds to high accuracy, the average
of (\ref{B2}) over noise realizations is
\begin{equation}
\langle Y_i \rangle = s_i(1 - \sum_k \beta^k_i) , \label{B3}
\end{equation}
since the `template' channels contain primordial CMB signal. If the 
coefficients  $\pmb{$\beta$}$ are independent of the signal, the covariance matrix  
$\langle Y_i Y_j \rangle$ is
\begin{equation}
\langle Y_iY_j \rangle \equiv {\bf C_Y} =  S_{ij}(1 - \sum_k \beta^k_i)(1 - \sum_k \beta^k_j) + N_{ij} + 
N^k_{ij}\beta^k_i\beta^k_j. \label{B4}
\end{equation}
The coefficients $\pmb{$\beta$}$ \ are found by solving
\begin{equation}
 \pmb{$\beta$}  = ({\bf F}^T{\bf C_Y(\pmb{$\beta$})}^{-1} {\bf F})^{-1}( {\bf F}^T {\bf C}^{-1}  {\bf x}), \label{B5}
\end{equation}
where ${\bf F}$ is a `large' $3N_p \times N_F$ matrix with elements constructed from the template maps $F^k_i$
(see equation (26) of  $[13]$). Finally, the likelihood for estimating cosmological parameters is computed
from 
\begin{equation}
{\cal L} \propto {1 \over \sqrt{ \vert {\bf C_Y} \vert}} {\rm exp} \left 
( - {1 \over 2} {\bf Y}^T{\bf C_Y}^{-1} {\bf Y} \right ) .   \label{B6}
\end{equation}

The motivation for the above procedure is discussed in detail in
$[13]$. It is equivalent to a Bayesian marginalization of foreground
templates subtracted from multifrequency maps in the limit that errors
in the foreground subtraction are small compared to the instrument
noise and cosmic variance. As demonstrated in $[13]$ this condition is
amply satisfied for the PSM if the highest and lowest polarized
\plancks channels are used as templates.

\subsection{Application to simulated \plancks data}

\begin{figure}[t]
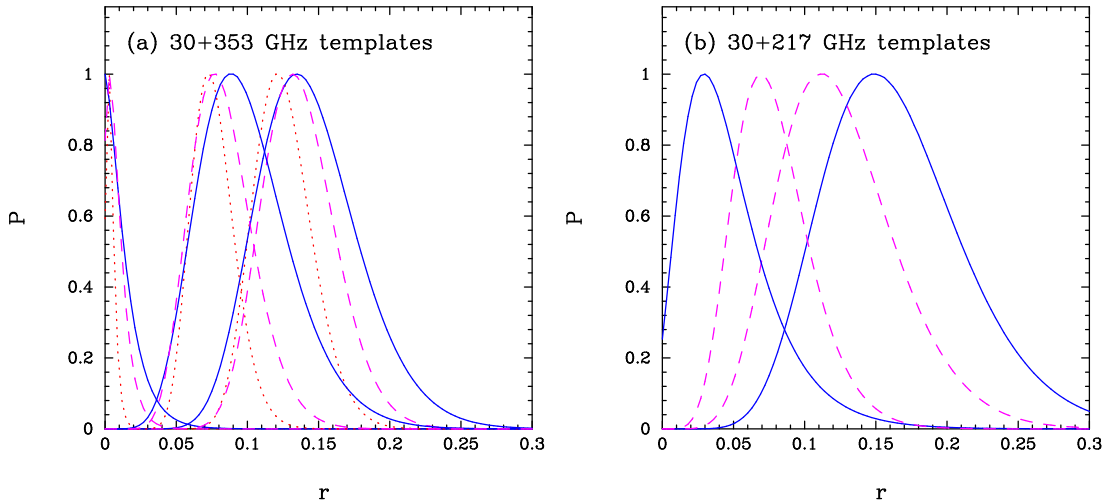


\vskip 3.0 truein

\includegraphics{pg_like1.ps}
\includegraphics{pg_like2.ps}

\caption {The likelihood (\ref{B6}) plotted as a function of the
  tensor-scalar ratio, $r$. Figure (a) to the left shows likelihoods
  for realizations with three values of the tensor-scalar ratio,
  $r=0$, $0.05$ and $0.1$ using the \plancks $30$ and $353$ GHz
  channels as templates. The solid blue lines show the likelihoods for
  the nominal \plancks mission of two sky surveys. The magenta dashed
  lines show the likelihoods for an extended \plancks mission of four
  sky surveys. The dotted red lines show the likelihoods for a mission
  with negligible noise.  Figure (b) to the right shows likelihoods
  for the simulations with $r=0.05$ and $r=0.1$, but now using the
  \plancks $30$ and $217$ GHz as templates. The colour coding is the
  same as in Figure \ref{figlike}a: solid blue lines show likelihoods
  for the nominal \plancks mission and dashed magenta lines show
  likelihoods for an extended \plancks mission.  In all cases, the
  polarization mask shown in Figure \ref{figmaps} has been applied to
  the simulated data.}

\label{figlike}

\end{figure}

For the simulations described here, we generated Gaussian realizations
of the primordial CMB at a Healpix $[15]$ resolution of $NSIDE=2048$
using the cosmological parameters for the concordance
$\Lambda$-dominated cold dark matter model determined from the WMAP 3
year data $[16]$. A Gaussian $B$-mode was added with a specified value
of $r$. Galactic polarized foregrounds from the PSM were added to the
primordial CMB at each of the \plancks polarized frequencies. Uniform
white noise, based on the detector sensitivities given in $[8]$ for
either the nominal mission lifetime (two sky surveys) or an extended
mission lifetime (four sky surveys) was added to the maps at each
frequency. These maps were then smoothed by a Gaussian of
$7^\circ$\ FWHM and repixelised to a Healpix resolution of $NSIDE=16$. An
internal polarization mask, constructed from the $217$ GHz PSM as
described in $[13]$ was applied to each of the low resolution maps
(see Figure \ref{figmaps} below). The mask is relatively conservative
and removes $37\%$ of the sky.

The main results of this paper are shown in Figure \ref{figlike}. This
shows the likelihood function (\ref{B6}) plotted as a function of $r$
for various noise levels and choice of templates\footnote{We assume
  that all cosmological parameters are fixed. This is a good
  approximation, since $r$ is weakly correlated with other
  cosmological parameters. We fix the tensor spectral index to $n_t =
  1$.}. For the results shown in Figure \ref{figlike}a, a `CMB
sensitive' map was constructed as an inverse noise variance weighted
sum of the four \plancks channels at $70$, $100$, $143$ and $217$
GHz. The $30$ GHz and $353$ GHz channels were used to define low and
high frequency templates. Likelihood distributions are plotted for
simulations generated with the same random numbers for three values of
the tensor-scalar ratio, $r=0$, $0.05$ and $0.1$. The red dotted lines
show results for nearly noise free simulations (diagonal noise of
$0.10\;\mu$K is added to the $Q$ and $U$ maps to regularize the signal
covariance matrix ${\bf S}$) of the primordial CMB alone. The solid
blue lines show the distributions after template subtraction for
simulations of the nominal \plancks mission, and the dashed magenta
lines show equivalent results for an extended \plancks mission.  As
expected, the distributions for the extended \plancks mission are
significantly narrower than those for the nominal distribution.  With
an extended mission a tensor-scalar ratio of $r = 0.05$ is detectable
at high significance. If $r \ll 0.05$, the likelihood function for
the extended mission drops to $0.05$ of its peak value at $r \sim
0.028$. An extended \plancks mission is therefore competitive with the
most sensitive of the $B$-mode optimised ground based/sub-orbital
experiments.

The improved signal-to-noise of an extended \plancks mission is
illustrated visually in Figure \ref{figmaps}. The upper panel shows
the noise-free realizations of the primordial $B$-mode contribution to
the $Q$ and $U$ maps for $r=0.1$. The middle panel shows the
reconstructed $B$-mode maps for the nominal \plancks mission using the
$30$ and $353$ GHz channels as templates. The lowest panel shows
$B$-mode reconstructions for an extended \plancks mission. Figure 3
shows quadratic maximim likelihood estimates (QML) $[17]$ of the $E$
and $B$-mode power spectra for the simulations with $r=0.05$. As
expected, the error bars on the extended mission power spectra
are almost half those of the nominal mission for the noise dominated
multipoles at $\ell \simgt 15$, but the effects of improved 
signal-to-noise can be seen over the entire range of multipoles 
shown in the Figure.

\begin{figure}

\vskip 5.5 truein

\includegraphics{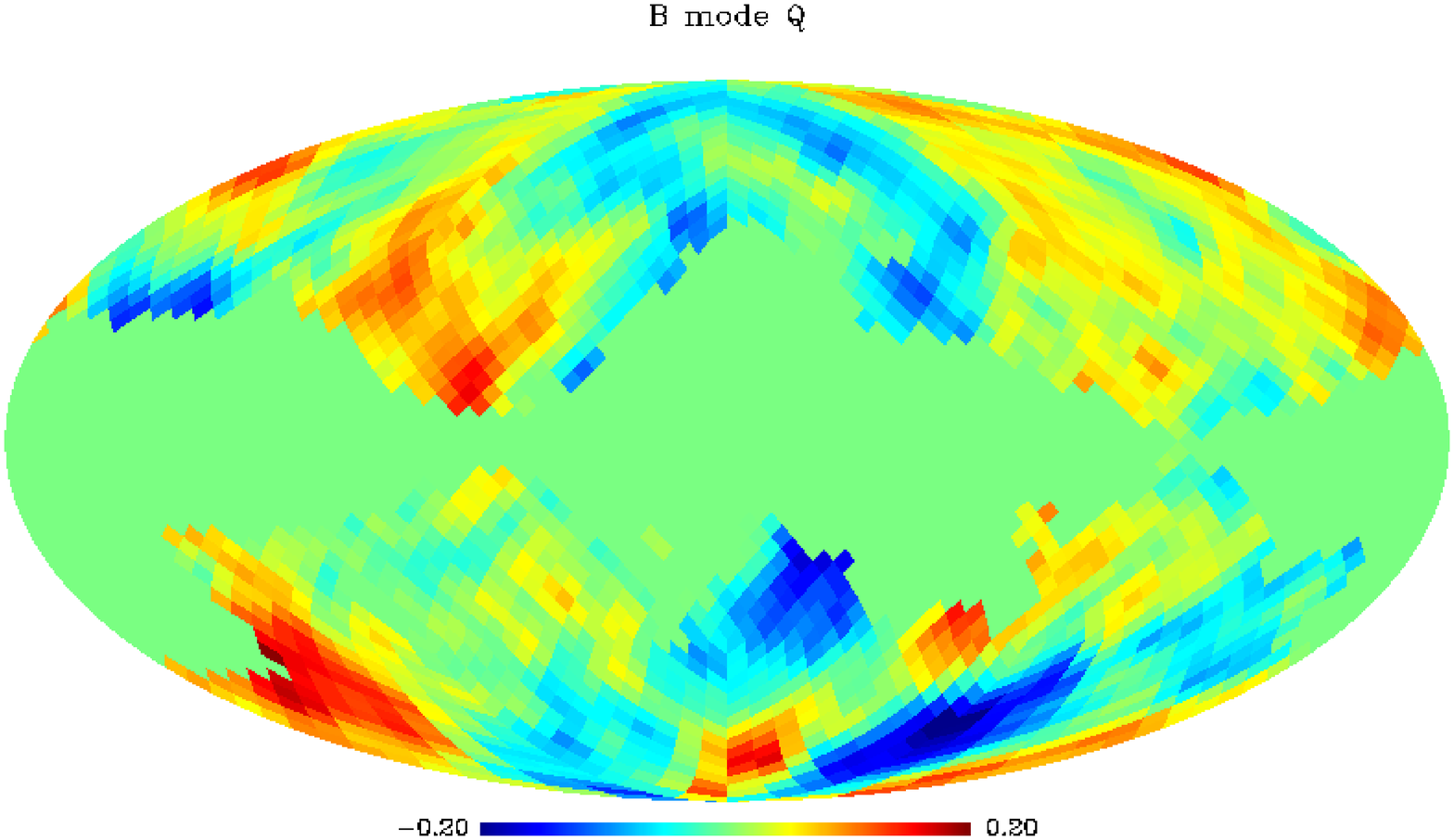}
\includegraphics{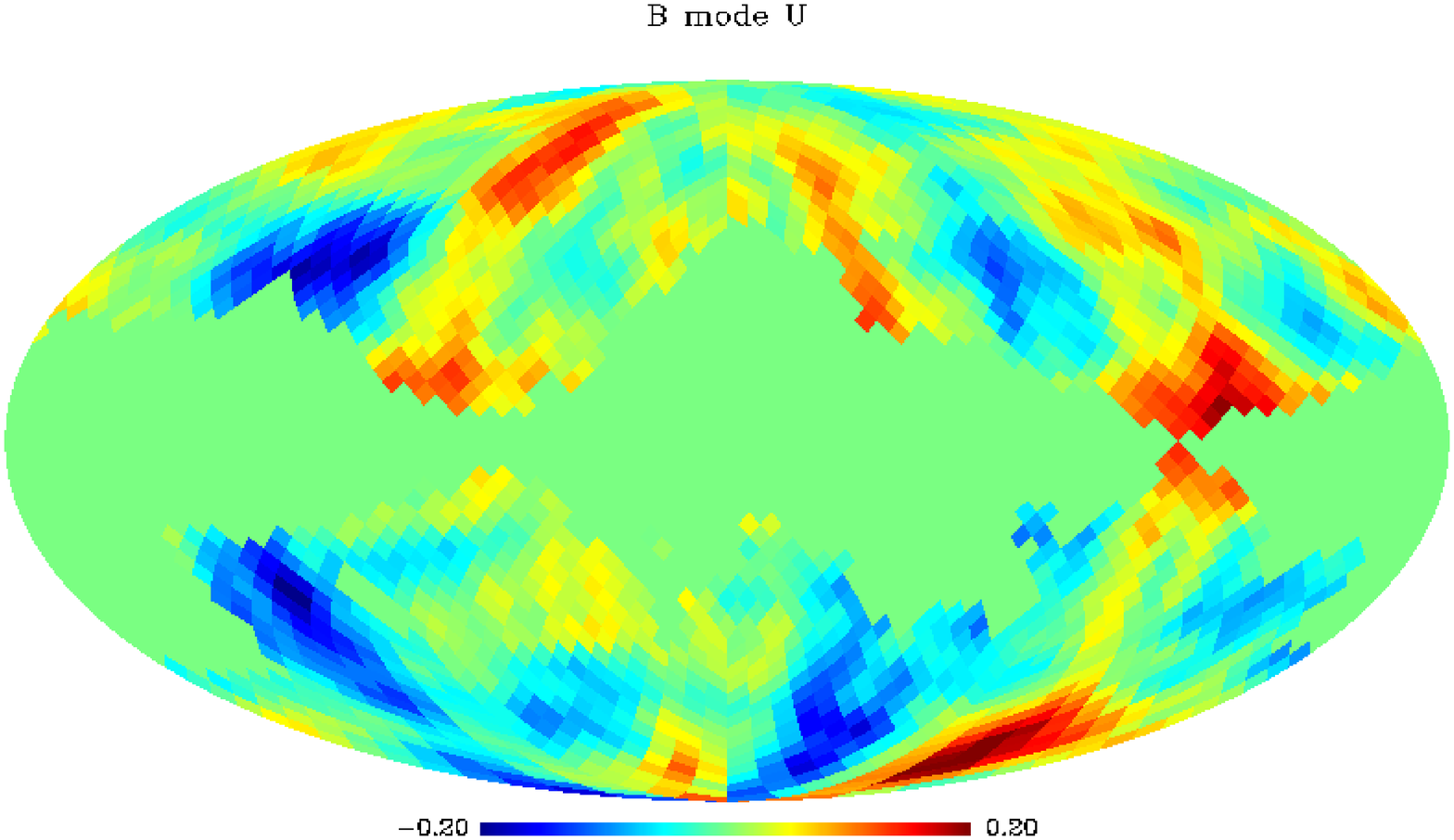}
\includegraphics{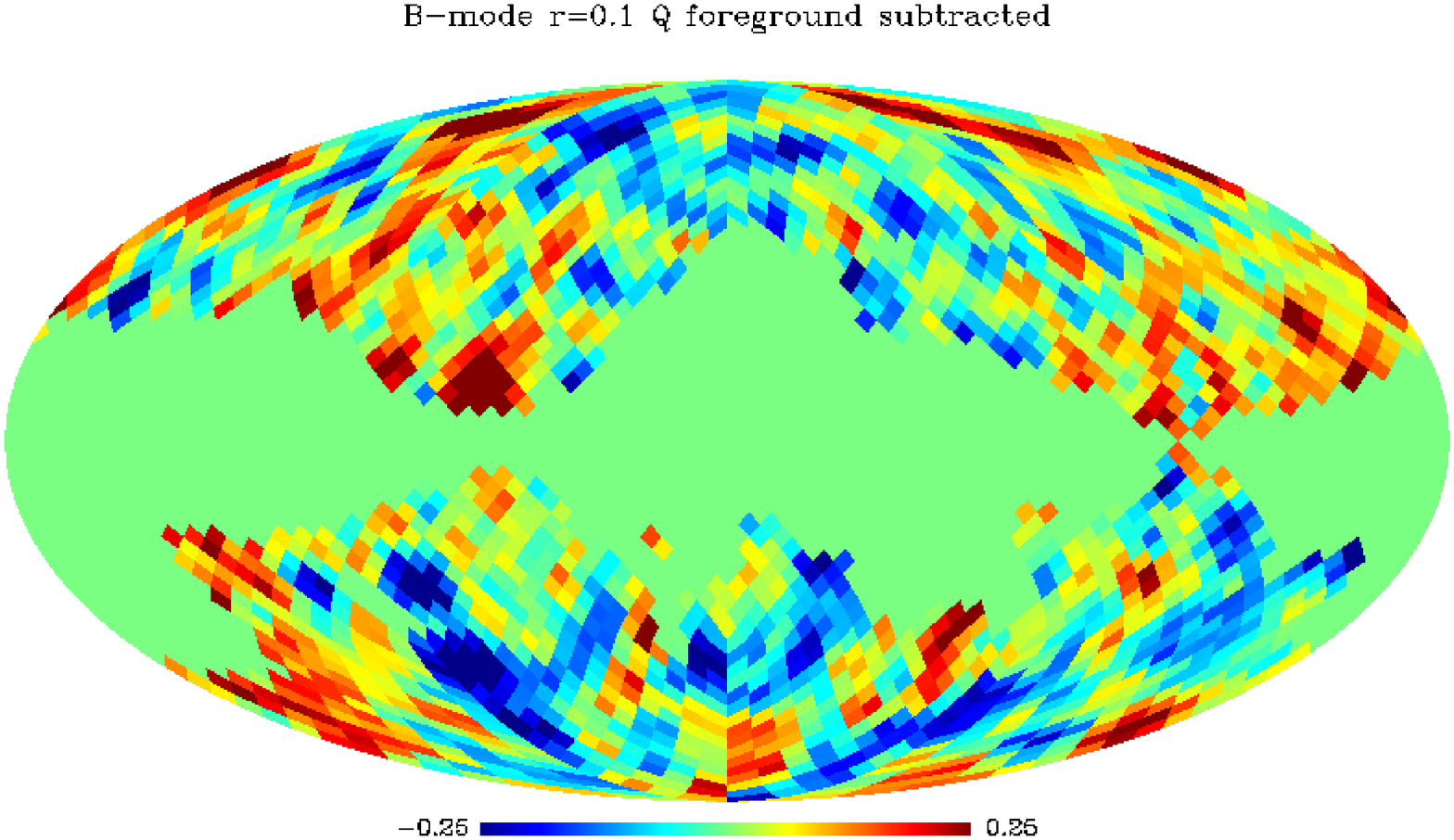}
\includegraphics{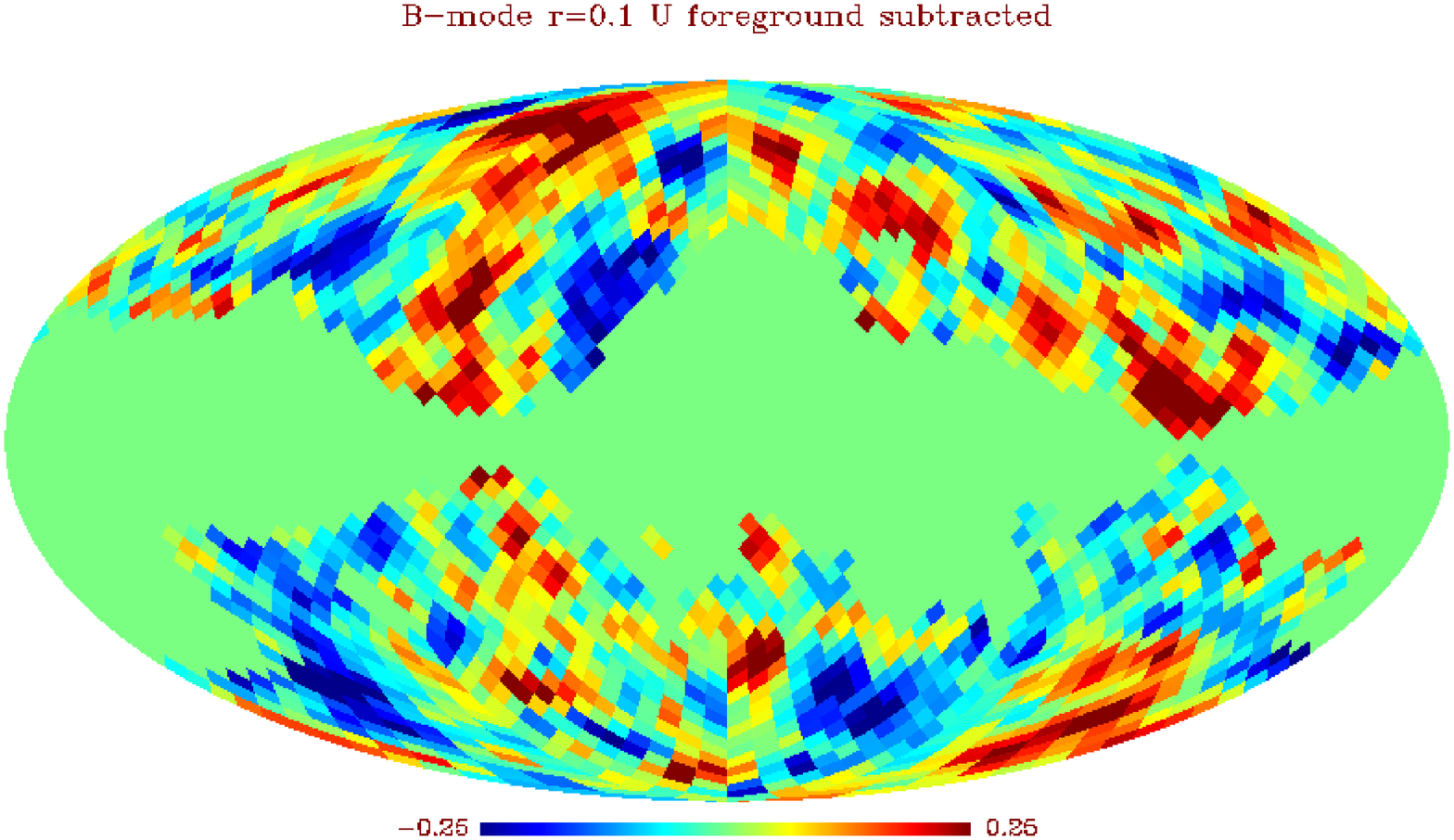}
\includegraphics{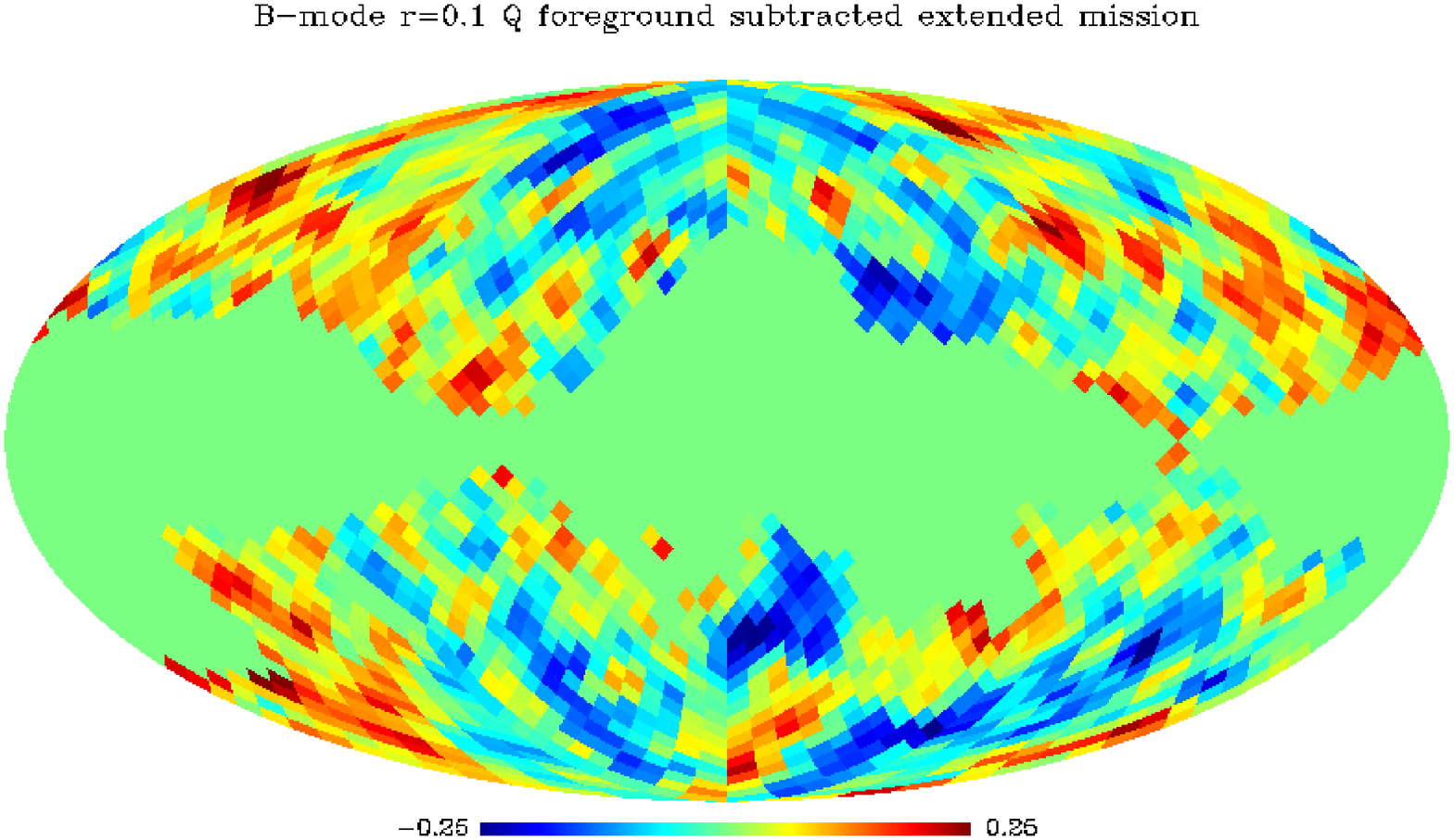}
\includegraphics{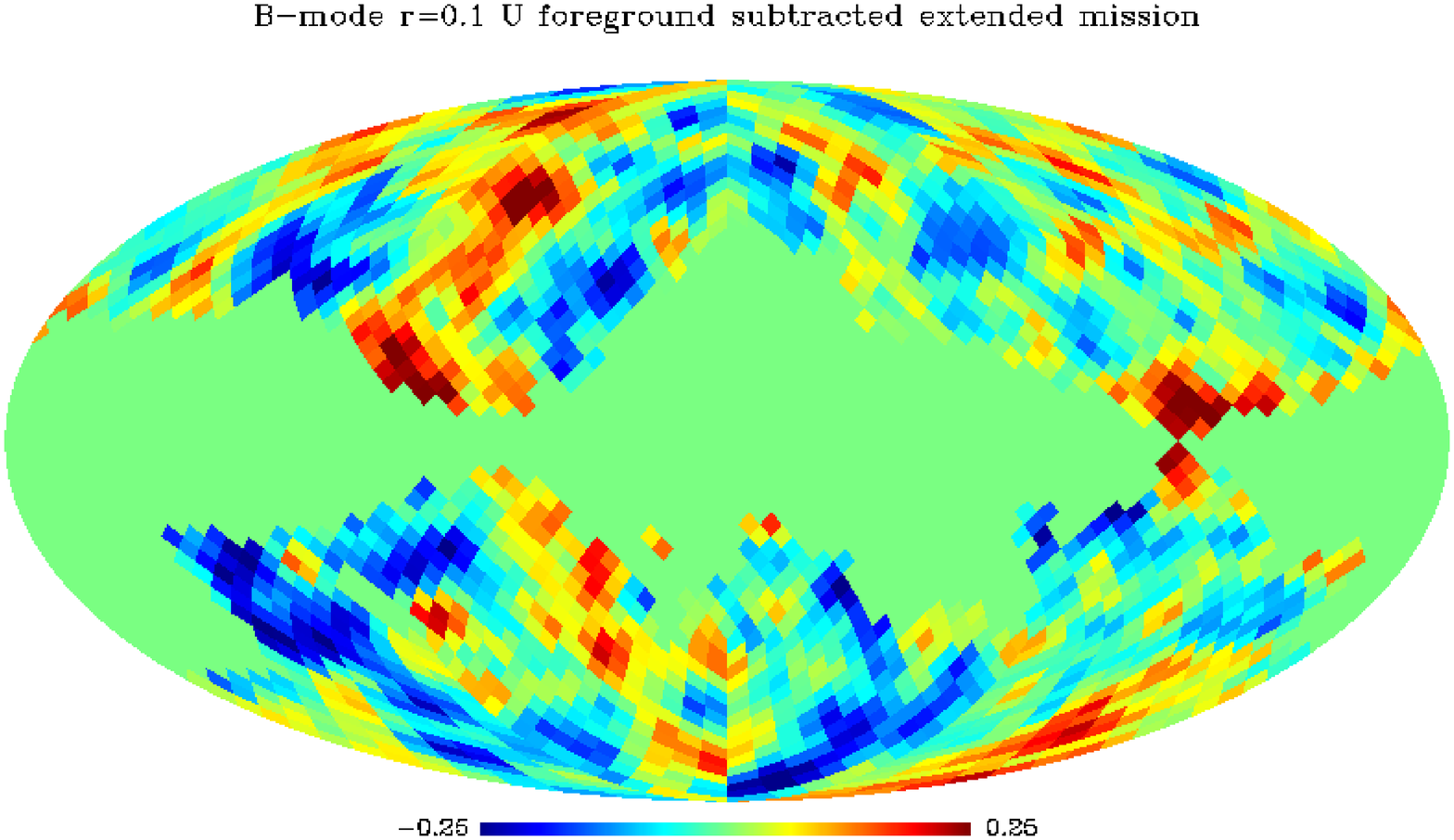}
\caption {$Q$ and $U$ maps smoothed with a Gaussian of FWHM=$7^\circ$ and pixelised
at a Healpix resolution of $NSIDE=16$. Upper panel shows the noise-free $B$-mode
contribution to the $Q$ and $U$ maps for a realization with $r=0.1$ for the region
outside an internally generated polarization mask. The middle panel shows the
foreground subtracted $B$-mode contribution for the nominal \plancks mission 
using the $30$ GHz and $353$ GHz channels as templates. Equivalent maps
for an extended \plancks mission are shown in the lower panel.}

\label{figmaps}

\end{figure}

\begin{figure}[t]
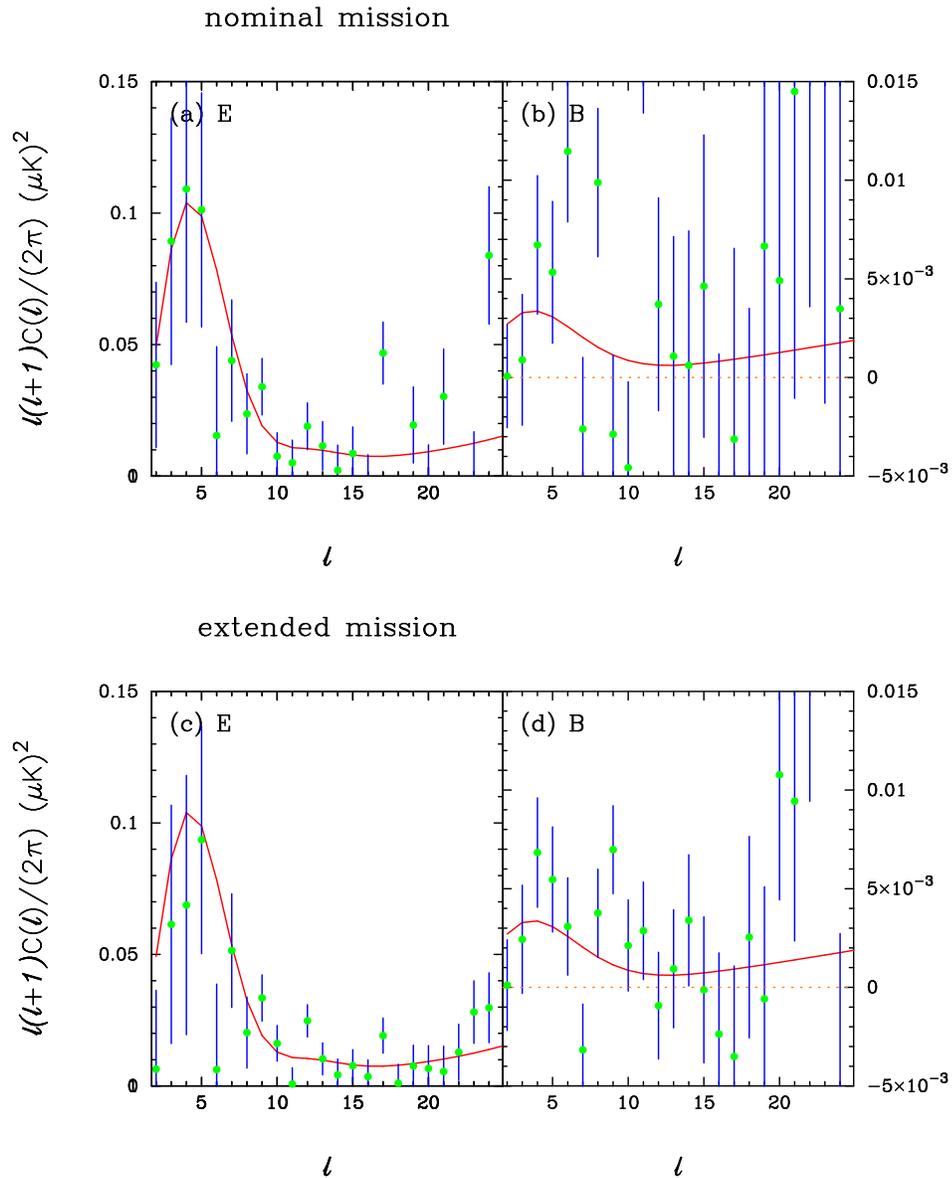


\vskip 6.5 truein

\includegraphics{pgqml0.05a.ps}
\includegraphics{pgqml0.05b.ps}

\caption {QML estimates of the $E$ and $B$-mode polarization spectra
  for the simulations with $r=0.05$.  Figures \ref{figqml}a and
  \ref{figqml}b show power spectra for the nominal \plancks
  mission. Figures \ref{figqml}c and \ref{figqml}d show power spectra
  for an extended \plancks mission. The error bars are computed from
  the diagonal components of the inverse of the QML Fisher matrix
  using the theoretical input spectra for $r=0.05$ (shown by the red
  lines).}

\label{figqml}

\end{figure}

An extended \plancks mission offers more than just reduced errors on
the polarization power spectra. The results described above were
derived using the specific choices of $30$ GHz and $353$ GHz \plancks
channels as templates.  WMAP has provided high quality maps of the
polarized Galactic emission over the whole sky at low frequencies
$[18]$ (particularly the $K$ and $Ka$ bands at $23$ GHz and $33$ GHz
respectively). In addition, the low-frequency instrument on \plancks
will provide all-sky maps at $30$, $44$ and $70$ GHz. These low
frequency maps will produce a wealth of data with which to constrain
polarized synchrotron emission. In contrast, the constraints on
polarized dust emission are substantially weaker and \plancks will
rely on data from its own high frequency polarization channels ($217$
and $353$ GHz) to provide dust templates over the whole sky. The
small number of polarized channels on \plancks limits the scope for
checking the accuracy of foreground separation, particularly at high
frequencies. The results described above and in $[13]$ show that if
polarized dust emission is as simple as assumed in the PSM,
then the \plancks $353$ GHz channel can provide an adequate dust
polarized template. However, even if this is the case, it will be
necessary to use the $217$ GHz channel as a template to demonstrate
consistency.

Figure \ref{figlike}b shows what happens if the $30$ and $217$ GHz
channels are used as templates. Here we have constructed inverse noise
variance weighted maps from the $70$, $100$ and $143$ GHz channels and
subtracted the $30$ and $217$ GHz channels as described in Section
2.1. (The simulations at each frequency are identical to those used to
construct Figure \ref{figlike}a).  Figure \ref{figlike}b shows the
likelihoods for the nominal and extended missions for the simulations
with $r=0.1$ and $r=0.05$. For the nominal mission, the likelihood
distributions are much broader than those shown in Figure \ref{figlike}a
Significantly, for the $r=0.05$ simulation, the nominal mission
sets only an upper limit on $r$. With an extended \plancks mission,
the distributions become much narrower and it is then possible to
detect $r=0.05$ at a high significance level. 

The $217$ GHz channel therefore becomes a useful dust template for an
extended \plancks mission, and as Figure \ref{figlike}b demonstrates
the improved constraints on $r$ from an extended mission are greater
than one might naively imagine from a $\sqrt 2$ improvement in the
detector noise. In both Figures \ref{figlike}a and \ref{figlike}b most
of the information on the primordial CMB anisotropies comes from the
$100$ and $143$ GHz \plancks bands, since these have high
sensitivity. To remove polarized dust emission, we  need to
subtract only about 5\% of the $353$ GHz maps, whereas we must subtract
about $26\%$ of the $217$ GHz maps. Although the $353$ GHz channel has
higher detector noise than the $217$ GHz channel, it has a higher
signal-to-noise for monitoring dust emission. Furthermore, by using
the $217$ GHz channel as a dust template, one can see from equations
(\ref{B3}) and (\ref{B4}) that a significant fraction of the primordial signal is 
subtracted from the data vector ${\bf Y}$, accentuating the impact of
detector noise. For the nominal \plancks mission these effects
significantly degrade \planck's sensitivity to inflationary $B$-modes if the
$217$ GHz channel is used as a dust template. However, with an extended
\plancks mission, the situation is substantially improved and it is
then feasible to use both the $217$ and $353$ GHz channels to subtract
polarized foregrounds thereby increasing the scope for redundancy
checks. The ability to perform such checks will be especially important
should \plancks reveal any evidence for a primordial $B$-mode
anisotropy.

\section{Conclusions}

Until a `CMBpol' satellite $[7]$ is launched, \plancks is
the only experiment capable of measuring primordial $B$-modes at
low multipoles. It is therefore important to assess carefully
\planck's capability for detecting $B$-modes and to consider how the
constraints on $B$-mode anisotropies improve if the \plancks mission
is extended to at least four sky surveys. This has been the goal of
this paper.

Any small primordial $B$-mode signal must be extracted from the
dominant polarized Galactic emission. Nevertheless, using the Planck
Sky Model, we have shown that an extended \plancks mission can set
strong constraints. An extended \plancks mission can set a $\sim 95\%$
upper limit of $r \approx 0.03$ and can detect a $B$-mode with
$r=0.05$ at a high significance level. Furthermore, with an extended
\plancks mission it is feasible to use both the $353$ and $217$ GHz
frequency bands to form dust templates suitable for probing
tensor-scalar ratios of $r \sim 0.05$.  This increases the scope for
testing the accuracy of Galactic foreground subtraction using \plancks
data, and also improves the prospects for dealing with polarized
foregrounds (particulary at high frequencies) should they turn out to
be more complicated than assumed in the PSM. We conclude that an
extended \plancks mission has comparable potential for primordial
$B$-mode detection as the most sensitive of the ground
based/sub-orbital experiments $[6]$ planned for the next few
years. Furthermore, \plancks is complementary to such experiments
because it is the only experiment capable of detecting inflationary
$B$-modes at low multipoles.  If a $B$-mode with $r \sim 0.1$ exists,
as predicted in `high field' inflation models with a power-law
potential $[19]$, there is a realistic prospect that within a few
years we might be able to measure the $B$-mode power spectrum over a
wide multipole range $2 \le \ell \simlt 200$.

Although we have focussed on $B$-mode detection in this paper, it is
worth mentioning that measurements of the $E$-mode power spectrum with
an extended \plancks mission will lead to interesting new science.
Accurate measurements of the $E$-mode spectrum will help to improve
our knowledge of the reionization history $[20]$. Furthermore, there
are indications from various analyses of the WMAP temperature data for
curious `anomalies' $[21]$, {\it e.g.}\ alignments of the low
multipoles, departures from statistical isotropy, and an unusual
`cold' spot. Many of the proposed physical explanations of these
apparent anomalies have polarization signatures at low multipoles
$[22]$ that are potentially observable by \planck.

\medskip

\noindent
{\it Acknowledgments:}
We thank STFC for financial report.  The authors acknowledge the use
of the Healpix package and of the Planck Sky Model developed by the
Component Separation Working group of the Planck Collaboration.

\end{document}